\documentclass[aps,pre,twocolumn,twoside,showpacs,floatfix,nofootinbib]{revtex4-1}
\usepackage{graphicx}

\begin{document}

\title{Heterogeneity shapes groups growth in social online communities}
\author{Przemys{\l}aw A. Grabowicz}
\email[Email: ]{pms@ifisc.uib-csic.es}
\affiliation{IFISC Instituto de F\'{\i}sica Interdisciplinar y Sistemas Complejos (CSIC-UIB), E07122 Palma de Mallorca, Spain}
\homepage{http://ifisc.uib-csic.es}
\author{V\'{\i}ctor M. Egu\'{\i}luz}
\affiliation{IFISC Instituto de F\'{\i}sica Interdisciplinar y Sistemas Complejos (CSIC-UIB), E07122 Palma de Mallorca, Spain}



\begin{abstract}
Many complex systems are characterized by broad distributions capturing, for example, the size of firms, the population of cities or the degree distribution of complex networks. Typically this feature is explained by means of a preferential growth mechanism. Although heterogeneity is expected to play a role in the evolution it is usually not considered in the modeling probably due to a lack of empirical evidence on how it is distributed. We characterize the intrinsic heterogeneity of groups in an online community and then show that together with a simple linear growth and an inhomogeneous birth rate it explains the broad distribution of group members.
\end{abstract}

\maketitle

\section{Introduction}
Many complex systems are characterized by \mbox{heavy-tailed} distributions, {\it e.g.}, Zipf's law (originally used to describe the frequency of words~\cite{Zipf1949Human}), Pareto's law (originally describing the wealth of nations~\cite{Pareto1964Cours}), and more recently scale-free topologies (capturing the degree distribution of complex networks~\cite{Barabasi1999Emergence})~\cite{Newman2005Power,Saichev2009Theory}. This property is typically perceived as a symptom of the rich-gets-richer principle, and models implementing some degree of preferential growth are usually the first approach to explain heavy-tailed distributions~\cite{Leskovec2008Microscopic,Mislove2008Growth,Eisenberg2003Preferential,Yamasaki2006Preferential,Simon1955class,Barabasi1999Emergence, Barabasi1999Mean-field, Huberman1999Internet:,Dorogovtsev2000Structure,Bornholdt2001World,Maruvka2011The,Hernandez2006Clone}.
In line with the rich-gets-richer principle, the Gibrat's law suggests that the expected growth of a firm, a city or social activity is proportional to its size~\cite{Gibrat1931Les,Gabaix1999Zipf,Rozenfeld2008Laws,Rybski2009Scaling}. However, in general, less attention has been devoted to the time evolution of complex systems probably due to the lack of empirical data along time (for some exceptions see \cite{Saichev2009Theory,Barabasi02,Palla2007Quantifying,Tessone10}). In many network growth models the time unit is mapped to the number of new arriving elements, which makes it difficult to compare the results with real data. Moreover, many models assume that the elements are born identical leading to correlations between age and frequency (of words, wealth, degree or size) which are not fully supported by empirical observations~\cite{Adamic2000Power-Law}. In many real systems, especially in social systems, individuals or elements are very diverse. In this direction, some models incorporating heterogeneity in the form of fitness, hidden variables or ranking have been proposed~\cite{Caldarelli2002Scale-Free,Soderberg2002General,Boguna2003Class,Fortunato2006Scale-Free,Ratkiewicz2010Characterizing}. However, there is rather little empirical work showing how intrinsic heterogeneity is distributed and its role in complex system growth~\cite{Garlaschelli2004Fitness-Dependent,DeMasi2006Fitness}. Based on data collected on a daily basis on the time evolution of an online social system we will characterize the heterogeneity of the groups and identify the heterogeneity and the distributed birth dates as key players explaining the heavy-tailed distribution of group sizes and the apparent proportional growth of groups to their size.

We study an online community called Flickr~\cite{Flickr}, where members can create and join groups. The groups in Flickr are mainly used to collaboratively post photos associated with the theme of the group. We will consider each group as an element of the system characterized by the number of members belonging to the group (group size). We have collected two datasets containing in total over 260,000 \mbox{member-created} groups in Flickr, which accounted for over 65\% of all public groups existing in Flickr. The first dataset has high temporal resolution and a wide time window. It contains 9,503 groups tracked for 350 days, between June 5, 2008 and May 20, 2009, by the publicly accessible external service called GroupTrackr~\cite{GroupTrackr}. The service tracked on a daily basis the number of members of the groups. The second dataset has shorter time window and minimal temporal resolution, but it covers a larger number of groups. It contains over 260,000 public groups for which we gathered information on the number of members, collected in two snapshots on December 18, 2009 and January 29, 2010. For these groups we also gathered estimated information on their birth date. As an estimation of the group birth date we consider the time when the first photo was posted to the group pool, as the first photo is normally posted to the pool soon after the group creation. The oldest groups in our dataset date back to July 16, 2004.

\begin{figure*}
\includegraphics[width=170mm]{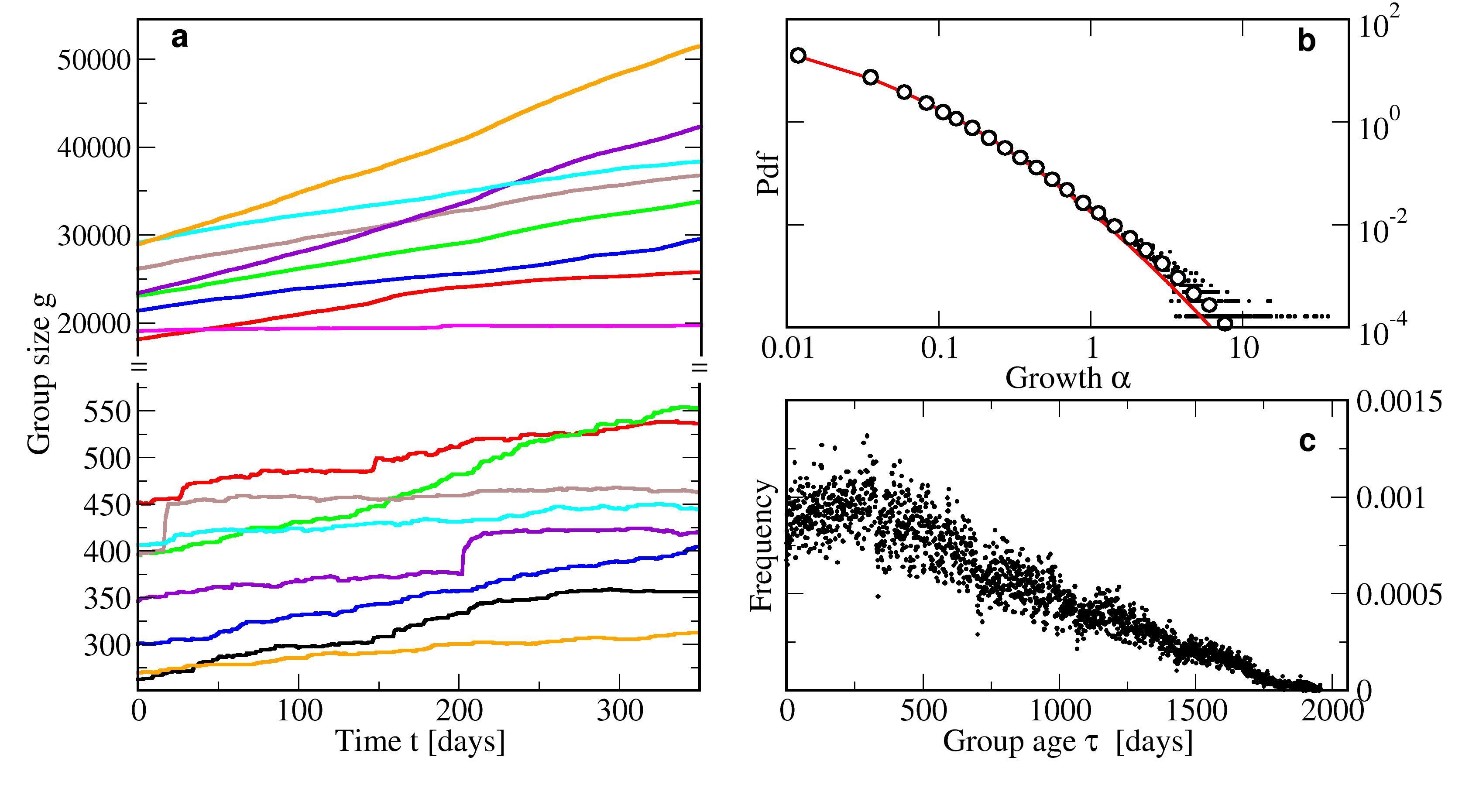}
\caption[ ]{(Color online) Characterizing the time evolution of online groups. (a) Time evolution of the group size for a representative sample of small and large groups. (b) Distributions of groups' growth $\alpha$ (open circles) with fitted log-normal distribution (line). The growth per day $\alpha$ is estimated based on growth over 6 weeks. (c) Distribution of group ages.}
\label{fig_lognormals}
\label{fig_parameters}
\label{fig_ages}
\label{fig_lineargrowth}
\end{figure*}

\section{Groups' growth in Flickr}
We first analyze the time evolution of groups. In Fig.~\ref{fig_lineargrowth}a we show how typical groups grow in number of members on a daily basis during the period of one year. As a first approach, a linear growth captures the individual trend (despite evident deviations in the form of sudden jumps). We have performed linear regression of time evolution of sizes of 9,503 groups over the period of almost one year. For about half of these groups the coefficient of determination $R^2$ has a value over 0.95, and more than $80\%$ of the groups larger than 1000 has $R^2$ higher than 0.95. The difference comes from the fact that the larger groups are affected less by the fluctuations of size. Aggregated residual plots do not show any clear trend deviating from our linear model. The time series cover considerable part of the average lifespan of the groups. Thus, we consider that groups grow linearly in time, the size $g_i$ of the group $i$ evolves as
\begin{equation}
g_i=1+\alpha_i(t-t_i^0)=1+\alpha_i\tau_i~,
\label{eq_alphatau}
\end{equation}
where $\alpha_i$ is the growth per unit of time, $t_i^0$ is the birth date and $\tau_i$ is the current age of group $i$.
We estimate the two parameters for 260,000 groups. The growth $\alpha_i$ for each group $i$ is calculated as the change of its size during 6 weeks, per day. A log-normal distribution provides the best fit to the distribution of growth values $\alpha$ (Fig.~\ref{fig_lognormals}b) with average $\mu=\overline{\ln \alpha} = -3.62$ and standard deviation $\sigma=1.57$. Finally, we estimated the current ages of all groups, finding that the number of groups created daily has been growing (almost linearly) in time (Fig.~\ref{fig_ages}c).

\section{Linear growth model with heterogeneous birth and growth}
Based on those findings we propose a minimal model of the time evolution of group sizes in Flickr, a linear growth model with heterogeneous birth and growth, which in short we will refer as the heterogeneous linear growth model. The model proceeds as follows, at each time step $t$: (i) new groups are created in the system. The number of groups created in each time step increases linearly with $t$. Each newly created group $i$ starts with one member and it is assigned its own growth value $\alpha_i$, drawn from a log-normal distribution. Growth value $\alpha_i$ remains unchanged for the simulation time; (ii) the size of each group $i$ is increased by $\alpha_i$.

\begin{figure}
\includegraphics[width=88mm]{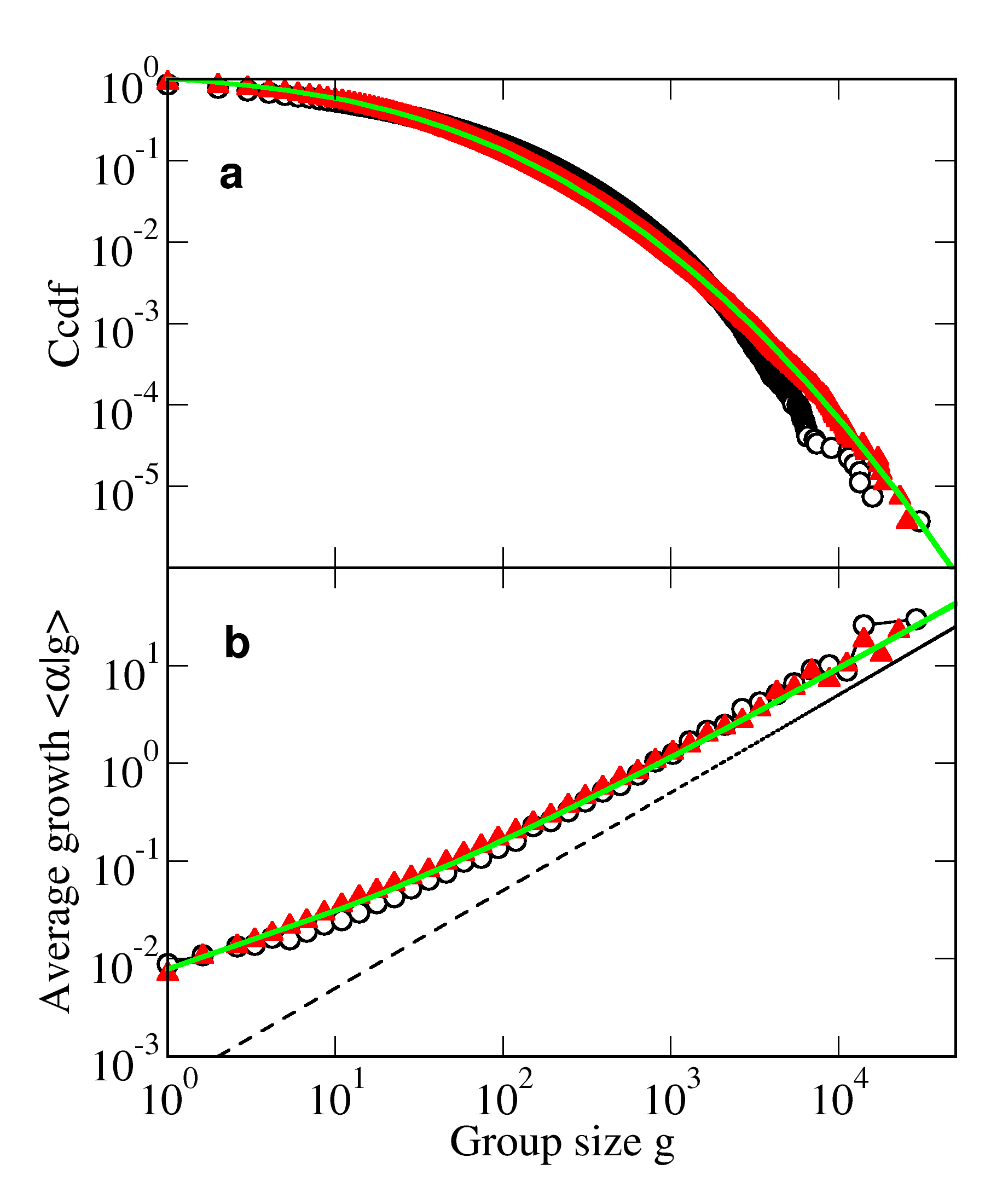}
\caption[ ]{(Color online) The heterogeneous linear growth model {\it vs.} real data. (a) Complementary cumulative distribution function of groups sizes for the real data (circles), the heterogeneous linear growth model (filled triangles) and its analytical solution (solid line). (b) Average daily growth as a function of the initial size of the groups, estimated for the period of 6 weeks and averaged over all groups of a given initial size, for:  the real data (circles), the model (triangles) and its numerical solution (line). The dashed line corresponds to the linear behavior $\langle \alpha | g\rangle \sim g$.}
\label{fig_gdistro}
\label{fig_avggrowth}
\end{figure}

We have run numerical simulations of the heterogeneous linear growth model where each time step of the simulation corresponds to a single day. We have simulated 1959 days in Flickr, from the moment when the first group from our dataset appeared. As a result of the numerical simulations we obtain the daily evolution of the sizes of over 260,000 artificial groups. The distribution of the final sizes of the groups reproduces with a good agreement the observed distribution (Fig.~\ref{fig_gdistro}a). As it can be seen from Fig.~\ref{fig_gdistro}a there is a small divergence for large group sizes, which could be explained by the deviations --mostly for small groups-- from the linear growth assumption. First, the strong fluctuations of the time evolution of group sizes of the small groups (see the jumps in Fig.~\ref{fig_lineargrowth}) lead to a larger 'apparent' growth than the real one, therefore leading to an over-estimation of their growth $\alpha$ and, as a consequence, the model displays a larger number of big groups than in the real system.

The average growth of groups of the same size, $\langle \alpha | g\rangle$, shows that bigger groups grow faster (Fig.~\ref{fig_avggrowth}b) both for the real data and the model in accordance with the Gibrat's law: $\langle \alpha | g \rangle \propto  g $. This result is obtained even though the microscopic rules of the model do not implement the \mbox{rich-gets-richer} principle. The average growth is an average over all groups of a given size, each of them growing linearly. Due to the heterogeneity and the linear growth, at a given time larger groups consist of old groups that grow slowly and younger groups that grow faster. Thus, the observation of preferential growth for groups of the same size does not reflect in this case an underlying rich-gets-richer principle, but it is a consequence of the competition of groups with different growth values and ages.

The statistical properties of the model can be estimated analytically. From the definition, the average growth of groups of the same size is given by:
\begin{equation}
\langle\alpha|g\rangle = \frac{ \int_{(g-1)/t} ^{\infty} \alpha p_{\alpha g}(\alpha,g)d\alpha } { \int_{(g-1)/t} ^{\infty} p_{\alpha g}(\alpha,g)d\alpha }~,
\label{ave}
\end{equation}
where  $p_{\alpha g}(\alpha,g)$ is the joint probability of having a group of size $g$ and growth rate $\alpha$, and $\int p_{\alpha g}(\alpha,g) d\alpha dg=1$. The lower limit of the integral is given by Eq.~(\ref{eq_alphatau}) and depends on $g$, and the maximum value of $\tau$ is limited to $t$, if the first group was created at time $t=0$. We transform Eq.~(\ref{ave}) replacing the joint probability $p_{\alpha g}(\alpha(g,g))$ by $p_{\alpha \tau}(\alpha(g,\tau))$ and making the assumption that $\tau$ and $\alpha$ are independent random variables:
\begin{eqnarray}
\langle\alpha|g\rangle & = & \frac
{ \int_{(g-1)/t} ^{\infty} \alpha p_{\alpha}(\alpha)p_{\tau}(\tau(\alpha,g)) \frac{\partial \tau}{\partial g} d\alpha }
{ \int_{(g-1)/t} ^{\infty} p_{\alpha}(\alpha)p_{\tau}(\tau(\alpha,g)) \frac{\partial \tau}{\partial g} d\alpha }.
\label{eq_def}
\end{eqnarray}
The numerical solution of Eq.~(\ref{eq_def}) for log-normal $p_{\alpha}$ and linear $p_{\tau}$ is plotted in Fig.~\ref{fig_avggrowth}b. Similarly the distribution of group sizes:
\begin{eqnarray}
p_g(g) & = & \int_{0} ^{\tau_{max}} p_{g \tau}(g,\tau) d\tau \\
       & = & \int_{0} ^{\tau_{max}} p_{\alpha}(\alpha(g,\tau))p_{\tau}(\tau) \frac{\partial \alpha}{\partial g} d\tau~, \label{eq_g}
\end{eqnarray}
is plotted in Fig.~\ref{fig_avggrowth}a. As one can see the solutions for both the average growth and the size distribution are in good correspondence with the results of numerical simulations, which indicates that the assumptions of independent random variables and linear growth are reasonable.\footnote{Equations~(\ref{eq_def}) and~(\ref{eq_g}) are easy to solve if $\alpha$ and $\tau$ are independent random variables and $p_{\alpha}$ is a power-law distribution. In such a case one can show that $\langle \alpha | g \rangle \propto g $ and that $p_g(g)$ is a power-law as well.}

\section{Heterogeneity vs. preferential growth}
\begin{figure}
\includegraphics[width=88mm]{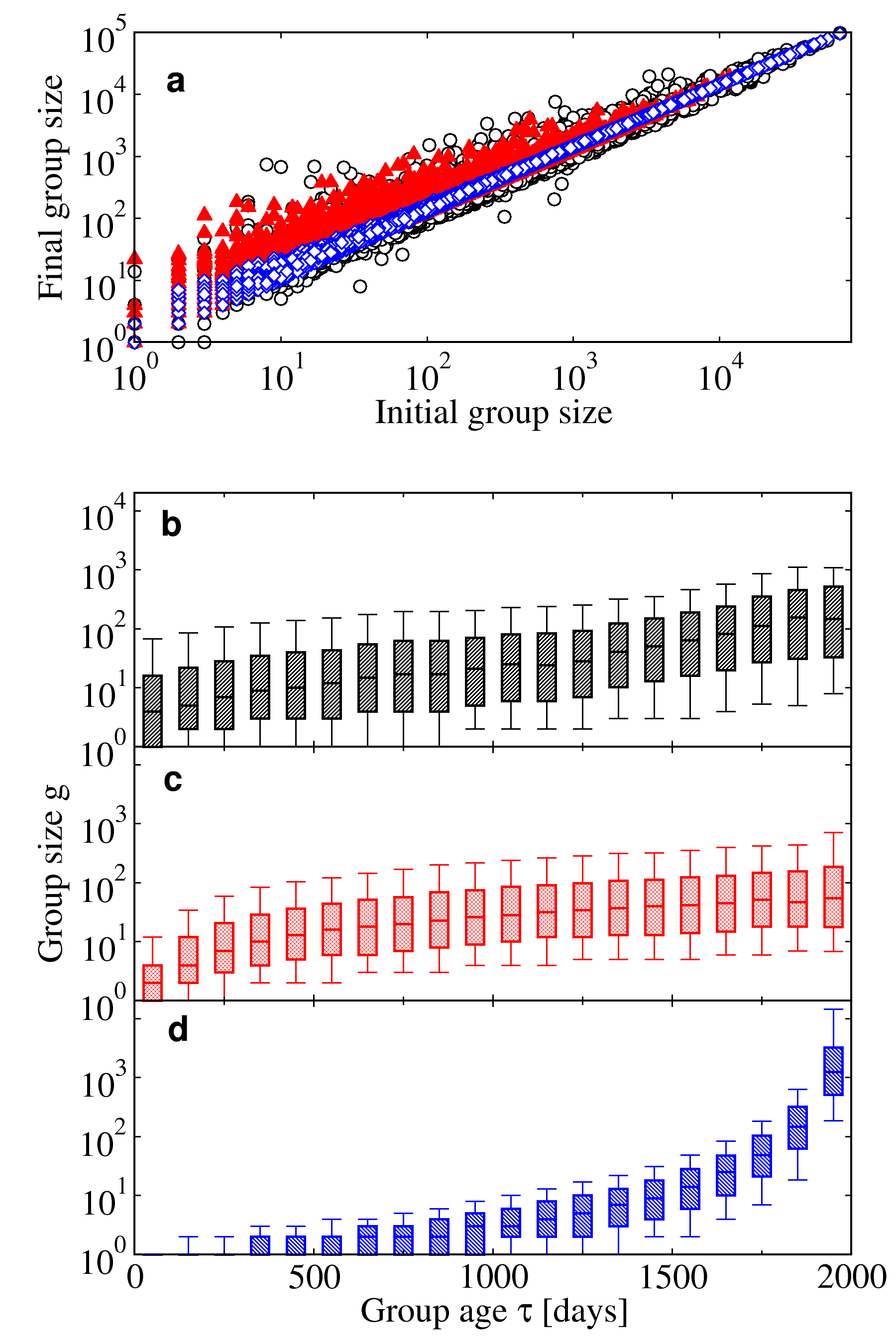}
\caption[ ]{(Color online) Comparison of Simon and heterogeneous linear growth model {\it vs.} real data. (a) Initial and final group sizes over a period of 350 days for the real data (circles), the heterogeneous linear growth models (filled triangles) and Simon model (diamonds). Each point represents a single group, there are 9,503 points plotted for each set of points. (b-d) Box plots with whiskers at 9th/91st percentile of final size of groups as a function of their age at the time of the measurement for 260,000 groups for (b) the real data, (c) the heterogeneous linear growth model, (d) the Simon model.}
\label{fig_modelscomparison}
\label{fig_gage123}
\end{figure}

We have shown that the heterogeneous linear growth model captures the statistical properties which commonly are attributed to the preferential growth mechanism. Thanks to the intrinsic heterogeneity, different growth patterns are permitted, even if groups have the same number of members at any point in time. One can see an example of this in Fig.~\ref{fig_lineargrowth}a, where group sizes are crossing themselves in time, though they continue to grow as they grew before the crossing. To make a direct comparison between the two mechanisms, heterogeneity {\it vs.} preferential growth, we consider the Simon model~\cite{Simon1955class}. The Simon model has been originally proposed to explain the distribution of words' frequency in a written text. At every time step, a word is added to the text: with a given probability $q$ it is a new word; otherwise, the word is chosen at random from the text, so the words which appear more frequently are chosen more often. We have adapted the Simon model to our system. We have set the parameters to obtain the same total number of groups and members as in the real case; also the number of new groups created in the system in each time step of the Simon model grows linearly, to isolate the effect of the heterogeneity. First, in the Simon model the final size of groups is heavily determined by their initial size measured one year before (Fig.~\ref{fig_modelscomparison}a), thus there is little heterogeneity among the groups, in contrast to the heterogeneous linear growth model which displays a degree of heterogeneity similar to the one of real groups. Second, for the Simon model the correlation of size and age is strong, while it is weak for real groups and the heterogeneous linear growth model (Figs.~\ref{fig_gage123}b-d)\footnote{In the heterogeneous linear growth model the average size of groups of given age is $\langle g|\tau\rangle = 1+\tau \exp{(\mu+\frac{\sigma^2}{2})}$, where $\mu$ and $\sigma$ are parameters of the lognormal distribution. In the Simon model, it is given by $\langle g|\tau\rangle = ( \frac { 2 + m T^2 } { 2 + m ( T - \tau )^2 } )^{1-q}$, where $T$ is the age of the system, $m$ controls the number of new users introduced into the system in each time step ($mT$), and $q$ is the probability of new group creation within the model (in our case $T=1959$, $m=10$ and $q=0.014$).}. The wide spread of group sizes corresponds to the high heterogeneity of groups, which is not captured by the preferential growth model (as observed in other systems as, for instance, in the World Wide Web where the number of links to the page is not strongly correlated with age of the web page~\cite{Adamic2000Power-Law}).

\section{Discussion}
In summary, we have proposed a simple growth model of heterogeneous elements with associated growing counters, based on the findings for a social system in an online community. We found that the model captures many of the features of the real system of online groups, namely the \mbox{heavy-tailed} distribution of group sizes, the average growth proportional to the current size of groups and the weak correlation between the age and the size of groups. Furthermore we made a direct comparison of the heterogeneous linear growth model with a preferential growth model and showed the similarities and the differences between these models. In the heterogeneous linear growth model the \mbox{heavy-tailed} distribution of final sizes of elements does not emerge from the growth process itself \mbox{(e.g., rich-gets-richer principle)}, but from the intrinsic heterogeneity of elements which take part in this growth process. This certainly does not answer the question why some groups grow faster than the others, as we do not understand yet what factors influence the fitness of the groups. However it points out that it does not have to be due to the fact that one group is bigger than the other as in preferential attachment models. The simplicity of our approach suggests that the characterization of the heterogeneity may play an important role in understanding the origin of broad distributions and the time evolution of many real systems.

\section*{Acknowledgements}
We thank Dario Taraborelli for the access to GroupTrackr data and help with further data collection process. P.A.G. and V.M.E. acknowledge partial support from NEST program of the European Commission through PATRES project, and from MICINN (Spain) through projects FISICOS (FIS2007-60327) and MODASS (FIS2011-247852); P.A.G. acknowledges support from the JAEPredoc program of CSIC (Spain).

\end{document}